\magnification=1200

\nopagenumbers 
\parindent=12pt
\baselineskip=20pt
\hsize=15 truecm
\vsize=23 truecm
\hoffset=0.7 truecm
\voffset=1 truecm
\overfullrule=0pt
\input epsf
\centerline{\bf BEYOND LEADING LOGARITHM}\par
\centerline{\bf PARTON DISTRIBUTIONS IN THE PHOTON} \par
\vskip 5 truemm
\centerline{\bf Michel FONTANNAZ} \par
\vskip 2 truemm
\centerline{Laboratoire de Physique Th\'eorique et Hautes Energies
\footnote{*}{Laboratoire
associ\'e au Centre National de la Recherche Scientifique (URA 63)}} \par
\centerline{Universit\'e de Paris XI, b\^atiment 211, 91405 Orsay Cedex, 
France} \par

\vskip 3 truecm
\noindent \underbar{{\bf Abstract}} \par
I discuss Beyond Leading Logarithm parton distributions in the photon and 
study the constraints
put on the latter by data on $F_2^{\gamma}(x, Q^2)$, and on the jet production 
in photo-proton
and photon-photon collisions.
 \par

\vfill \supereject
\noindent {\bf 1. \underbar{Introduction}}
\vskip 5 mm
For the past fifteen years, the photon structure function $F_2^{\gamma}(x, 
Q^2)$,
measured in deep inelastic lepton-photon scattering (Fig. 1), has generated
considerable theoretical and experimental work [1], and, more recently, the 
beginning
of HERA has reactivated the interest in the quark and gluon distributions in 
real
photons. It is indeed expected that photoproduction experiments [2, 3] will 
allow to
measure these distributions with a good precision. Another interest of 
photoproduction
reactions consists in the coupling of the gluon (from the photon) to the hard
subprocesses (Fig. 2) which offers a direct determination of its distribution 
[4]~: in
$\gamma \gamma^*$ DIS experiments it is only through the evolution equations 
for
$F_2^{\gamma}$ that the gluon distribution may be determined.
$$ 
\epsfbox{fig1} 
$$
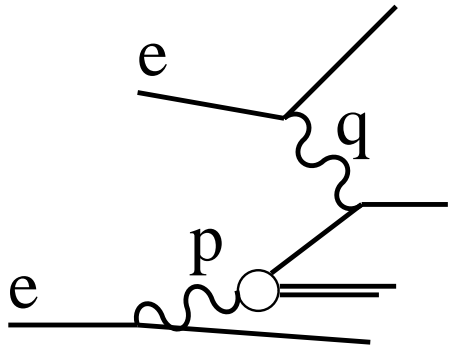
\centerline{Fig. 1 : Deep inelastic scattering on a real photon ($p^2 \simeq 
0)$.} \par
\vskip 3 mm
Jet production in $\gamma \gamma$ collisions also offers the possibility to 
measure the quark and
gluon distributions in real photons, and data from TRISTAN [5, 6] are very 
interesting. In this
reaction the parton distributions may doubly intervene and the cross-section 
is very sensitive to
the latter. The $x$-region probed at HERA ($x \sim .1$) and TRISTAN ($x \sim 
.4$) are complementary
whereas the $Q^2$-regions are very similar ($Q^2 \simeq p_{\bot}^2 =$ 25 
GeV$^2$). Therefore the
joint study of HERA and TRISTAN [7, 8] results should constrain rather tightly 
these parton
distributions.
$$
\epsfbox{fig2}
$$

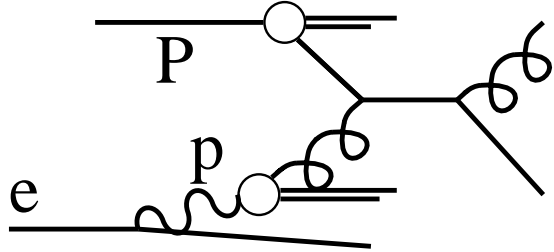
\noindent Fig. 2 : Jet photoproduction through the subprocess gluon + quark $
\to$ gluon + quark.

\par \vskip 3 mm
In this talk, I would like to discuss the photon structure function, stressing 
the importance of the
Higher Order QCD corrections, of the factorization scheme and of the problem 
of the non perturbative
input to $F_2^{\gamma}$. In particular I show how the non peturbative 
component must be modified in
order to take into account the specificity of the $\overline{MS}$ 
factorization scheme. \par

Then I will compare the theoretical predictions with experimental results on 
$F_2^{\gamma}$, and
discuss the possibility to obtain indications on the non perturbative input 
from data. Finally I
will consider jet production in $\gamma \gamma$ collisions and in 
photoproduction and the
constraints put by the data on the quark and gluon distributions. \par

\vskip 5 mm
\noindent {\bf 2. \underbar{The parton distributions in the photon}} \par
\vskip 5 mm
The parton contents of the photon can be measured in deep inelastic scat\-te
\-ring ex\-pe\-ri\-ments
in which the virtual photon $\gamma^{\ast}$ of momentum $q$ ($Q^2 = - q^2 >> 
\Lambda^2$) probes the
short distance behavior of the real photon $\gamma$ of momentum $p$ (Fig.~1). 
The structure
function $F_2^{\gamma}$ of this reaction is proportional, in the Leading 
Logarithm (LL) approximation,
to the quark distributions in the real photon
$$F_2^{\gamma}(x, Q^2) = x \sum_{f=1}^{n_f} e_f^2 \left ( q_{\gamma}^f(x, Q^2) 
+
\bar{q}_{\gamma}^f(x, Q^2) \right ) \ \ \ . \eqno(1)$$

\noindent  The sum in (1) runs over the quark flavors and $x = Q^2/2p.q$. \par 

It is instructive to consider the contribution to $F_2^{\gamma}$ of the lowest 
order diagrams of
Fig.~3. Contrarily to the case of a hadronic target, the lower part of the 
diagram is known~: it is
given by the coupling of photon to quark.
$$
\epsfbox{fig3}
$$
\noindent This contribution is therefore exactly calculable, with the 
following result
for a quark of charge $e_f$~:
$$F_2^{\gamma}(x, Q^2)/x = 3e_f^4 {\alpha \over \pi}  \left \{ \left ( x^2 + 
(1-x)^2 \right )
\ell n {Q^2 \over m_f^2} \right.$$
$$\left . + \left ( x^2 + (1-x)^2 \right ) \ell n {1-x \over x} +
8x(1-x) -1 \right \} . \eqno(2)$$

However our result (2) is not directly related to a physical process, because 
it depends on the
unknown quark mass $m_f$, used as a cut-off to regularize a logarithmic 
divergence. Actually this
perturbative approach is certainly not valid when the virtuality $|k|^2$ of 
the exchanged quark
becomes small. We then go into a non perturbative domain where we lack 
theoretical tools and we must
resort to models to describe non perturbative (NP) contributions to $F_2^{
\gamma}$. A popular model
is the ``Vector Meson Do\-mi\-nan\-ce Model'' (VDM) which consider that the 
real photon couple to
vector mesons. Therefore the real photon, besides a direct coupling to a $q 
\bar{q}$ pair,
has a VDM component which is also probed by the virtual photon. \par

The latter component contributes to $F_2^{\gamma}$ and must be added to 
expression (2). Keeping only
the term in (2) proportional to $\ell n \ Q^2/m_f^2$ (LL approximation), we 
write
$$F_2^{\gamma}(x, Q^2) = 3 e_f^4 {\alpha \over \pi} x \left ( x^2 + (1 - x)^2 
\right ) \ell n {Q^2
\over Q_0^2} + x \sum_{V=\rho, \omega , \phi} e_f^2 \left ( q_f^V(x) + 
\bar{q}_f^V(x) \right ) \ \ \
. \eqno(3)$$

\noindent The scale $Q_0^2$ is the value of $Q^2$ at which the perturbative 
approach is no more
valid. The perturbative contribution vanishes at $Q^2 = Q_0^2$ and $F_2^{
\gamma}$ is described only by
the non perturbative contribution $q_f^{NP}(x) = q_f^V(x)$ which describes the 
quark contents of
vector mesons. \par

We have to keep in mind that this way of treating the non perturbative part of 
$F_2^{\gamma}$ is
due to our lack of theoretical understanding of this contribution. There are 
other
approaches [9], especially that of Ref. [10] which takes into account the 
interaction between the
quarks and the gluon condensate. These different approaches must ultimately be 
compared with
experiment. \par

QCD corrections to the diagrams of Fig. 3 do not change the basic structure of 
expression (3). In
the LL approximation [11, 12], the perturbative quark distribution is given by 
the sum of ladder
diagrams (Fig. 4) (for simplicity we consider only one flavor),
$$
\epsfbox{fig4}
$$
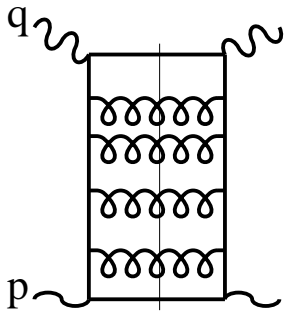
\centerline{Fig. 4 : Ladder diagram contribution to $F_2^{\gamma}$ (the thin 
line cuts final
partons).} \par
\vskip 3 mm 

\noindent  whereas the non perturbative part acquires a $Q^2$-dependence which 
is identical to that
of a quark distribution in a hadron. Thus the total quark and gluon 
distributions are given by (AN is
for anomalous, a designation introduced in Ref. 11 for the perturbative 
distribution) 
$$\eqalignno{
&q_{\gamma}(n,
Q^2) = q_{\gamma}^{AN}(n, Q^2) + q_{\gamma}^{NP}(n, Q^2) \equiv d_q(n, Q^2) \cr
&g_{\gamma}(n, Q^2) = g_{\gamma}^{AN}(n, Q^2) + g_{\gamma}^{NP}(n, Q^2) \equiv 
d_g(n, Q^2) &(4) \cr
}$$ 

\noindent  which verify the inhomogeneous equation ($i, j = q,\bar{q},g$)

$$Q^2 {\partial d_i(n, Q^2) \over \partial Q^2} = {\alpha \over 2 \pi} 
k^{(0)}_i(n) +
{\alpha_s(Q^2) \over 2 \pi} P_{ij}^{(0)}(n) d_j(n, Q^2) \ \ \ . \eqno(5)$$

\noindent  As in (3) we introduce the boundary condition [13] $Q_0^2$ in (4) 
so that
$q_{\gamma}^{AN}(n, Q_0^2)$ $g_{\gamma}^{AN}(n, Q_0^2)$ vanish when $Q^2 = 
Q_0^2$. $k^{(0)}_q(n) =
\int_s^1 dx \ x^{n-1} k^{(0)}_q(x)$ is the moment of the Altarelli-Parisi 
kernel des\-cri\-bing the
splitting of a photon into a $q \bar{q}$ pair (the bottom rung of the ladder) 
whereas $P_{ij}^{(0)}$
are the usual AP kernels. \par

The modifications of these LL results due to Higher Order (HO) QCD corrections 
[14, 15, 16] are
obtained by replacing the LL kernels of (5) by kernels involving HO 
contributions
$$\eqalignno{
k_i(n) & = {\alpha \over 2 \pi} k_i^{(0)}(n) + {\alpha \over 2 \pi} {\alpha_s 
\over 2 \pi}
k_i^{(1)}(n) + \cdots \cr
P_{ij}(n) & = {\alpha_s \over 2 \pi} P_{ij}^{(0)}(n) + \left ( {\alpha_s \over 
2 \pi} \right )^2
P_{ij}^{(1)}(n) + \cdots \ \ \ , &(6) \cr
}$$ 

\noindent  and by a modification of the expression of $F_2^{\gamma}$ in terms 
of parton
distributions (the gluon contribution is now explicitly written)
$$F_2^{\gamma}(n - 1, Q^2) = e_f^2 C_q(n, Q^2) \left ( q_{\gamma}(n, Q^2) + 
\bar{q}_{\gamma}(n,
Q^2) \right ) + C_g(n, Q^2) g_{\gamma}(n, Q^2) + C_{\gamma}(n) \eqno(7)$$

\noindent  where $C_{\gamma}$ is the ``direct term'', given by the part of (2) 
not proportional to
$\ell n {Q^2 \over m_f^2}$. $C_q$ and $C_g$ are the well-known Wilson 
coefficients which are
identical to those found in the case of a hadronic target. \par

A delicate point when working beyond the LL approximation is that of the 
factorization scheme. A
change in the factorization scheme is translated into a change in $k^{(1)}$ 
and $C_{\gamma}$
but in such a way that the physical quantity $F_2^{\gamma}$ remains unmodified 
(at order
$\alpha_s^0$). On the other hand $q_{\gamma}$ is not an invariant with respect 
to the factorization
scheme and a change in $k_i^{(1)}$ causes modifications in $q_{\gamma}^{AN}$ 
\underbar{and}
$q_{\gamma}^{NP}$. Therefore the separation (4) in a perturbative and a
 non perturbative part is \underbar{not} scheme invariant and the statement 
that $q_{\gamma}^{NP}$ can
be described by VDM has no meaning, unless one specifies in which 
factorization scheme it is valid.
\par

Within the $\overline{MS}$ scheme we obtain for $C_{\gamma}$ the expression 
given by the part of Eq.
(2) not proportional to $\ell n Q^2/m_f^2$. We observe that the $\ell n(1 - 
x)$ factor becomes very
large near the boundary of phase space and then it does not appear as a 
correction in Eq. (7) since
it may become numerically larger that the leading terms $q_{\gamma}$ which are 
enhanced by a factor
$1/\alpha_s$. In order to keep the concept of a perturbative expansion useful, 
it is proposed in
[17] to introduce the $DIS_{\gamma}$ factorization scheme where the choice
$$\left . C_{\gamma}(x)\right|_{DIS_{\gamma}} = 0 \eqno(8)$$

\noindent is made. The parton distributions thus defined satisfy an evolution 
equation of type (5,
6) with the inhomogeneous terms $k^{(1)}$ replaced by
$$\eqalignno{
&\left . {\alpha \over 2 \pi} k_q^{(1)} \right |_{DIS_{\gamma}} = {\alpha 
\over 2 \pi} k_q^{(1)} -
P_{qq}^{(0)} \ C_{\gamma}/2e_f^2 \cr 
&\left . {\alpha \over 2 \pi} k_g^{(1)} \right |_{DIS_{\gamma}} = {\alpha 
\over 2 \pi} k_g^{(1)} -
P_{gq}^{(0)} \ C_{\gamma}/e_f^2  &(9) \cr
}$$ 

\noindent as can be immediately derived by expressing the $\overline{MS}$ 
parton distributions
$q_{\gamma}$ in terms of their $DIS_{\gamma}$ counterparts
$$\eqalignno{
&\left . q_{\gamma} = q_{\gamma} \right |_{DIS_{\gamma}} - C_{\gamma}/2e_f^2 
\cr
&\left . g_{\gamma} = g_{\gamma} \right |_{DIS_{\gamma}} &(10) \cr
}$$

\noindent in Eq. (5, 6). Note that the homogeneous terms are not affected by 
this transformation.
It is shown in [17] that, in this convention, the leading logarithmic and the 
beyond leading
logarithmic parton distributions remain very close to each other over the all 
range in $x$. \par

We adopt here a different approach which also absorbs the troublesome 
``large'' $\ell n (1 - x)$
terms with the added advantages that the parton distributions we define are 
universal (i.e.
independent of the reference process) and obey the $\overline{MS}$ evolution 
equations, as all hadron
structure functions in practical use today. A careful analysis [18] of the box 
diagram (Fig. 3)
indeed shows that it is possible to define a factorization-scheme-invariant 
non perturbative input
$\bar{q}_{\gamma}^{NP}(n, Q^2)$, related to the non perturbative input defined 
in (4) by the relation
$$q_{\gamma}^{NP}(n, Q_0^2) = \bar{q}_{\gamma}^{NP}(n, Q_0^2) - C_0(n)/2e_f^2 
\eqno(11)$$

\noindent with $C_0(x)$ given in the $\overline{MS}$ scheme by
$$C_0(x) = 3e_f^4 {\alpha \over \pi} \left \{ \left ( x^2 + (1 - x)^2 \right ) 
\ell n (1 - x) + 2x(1
- x) \right \} \ \ \ . \eqno(12)$$

\noindent Actually all the scheme dependence of $q_{\gamma}^{NP}$ (Eq. 4) is 
contained in
$C_0(x)$~; $\bar{q}_{\gamma}^{NP}$ is factorization scheme independent and we 
use the Vector Meson
Dominance Model to describe it. \par

\vbox to 9.5 truecm {}
\baselineskip=14 pt
\noindent Fig. 5 : Predictions for ${\rm F}_2^{\gamma}$ compared
with AMY data [19]. Non perturbative input of Eq. 11 with $\bar{q}_{
\gamma}^{NP} = q_{\gamma}^{VDM}$
and $Q_0^2 = .25 \ {\rm GeV}^2$ (full curve)~; same input and $Q_0^2 = 1. \ {
\rm
GeV}^2$ (dotted curve)~; input of Eq. 11 with $C_0(n) = 0$ and $Q_0^2$ = .25 
GeV$^2$ (dashed curve).
\par \vskip 3 mm
\baselineskip = 20 pt
Let us now turn to a comparison between the theoretical calculations and the 
experimental results.
Details on the VDM input and on the treatment of the massive charm quark can 
be found in Ref. [18].
In Fig. 5, we see a prediction obtained with $Q_0^2 = .25$ GeV$^2$ ($\Lambda_{
\overline{MS}}$ = 200
MeV) compared with AMY data [19]. The sensitivity to the $C_0(x)$ term of Eq. 
(11) is also
shown, as well as the sensitivity to the value of $Q_0^2$. A similar 
comparison with JADE data
[20] is displayed in Fig.~6 on which we can observe the effect of the non 
perturbative input
described by VDM. \par

\vbox to 11 truecm {}
\baselineskip = 14 pt
\noindent Fig. 6 : JADE data [20]. Non perturbative input of Eq. 11 with $
\bar{q}_{\gamma}^{NP}
= q_{\gamma}^{VDM}$ and $Q_0^2$ = .5 GeV$^2$ (full curve). With $\bar{q}_{
\gamma}^{NP} = 0$ (dashed
curve). \par \vskip 3 mm \baselineskip = 20 pt

We clearly see from these comparisons that a satisfactory agreement is 
obtained between theory and
experiment but the accuracy of the data are not good enough to put constraints 
on the non
perturbative input and on the value of $Q_0^2$. Let us however notice that we 
compared with
large-$Q^2$ data corresponding to an important contribution of the anomalous 
distribution (Eq. 4)
which has a term proportional to $\ell n Q^2/\Lambda^2$. The low-$Q^2$ region 
emphasizes the role of
the non perturbative input as it can been observed in Fig. 7 which displays 
PLUTO [21] and TOPAZ [22]
data. The nice agreement of the theoretical prediction with these data shows 
that the non
perturbative input is reasonably described by VDM and a value $Q_0^2 = .5$ 
GeV$^2$. \par

\vbox to 8 truecm {}
\baselineskip = 14 pt
\noindent Fig. 7 : PLUTO data [21] and TOPAZ data [22]. Same as fig. 6.  

\par \vskip 3 mm \baselineskip = 20 pt

In this talk I only give comparisons between the distributions functions of 
Ref.~[18] and some data.
Two other Beyond Leading Logarithm (BLL) parametrizations of the parton 
distributions in the photon
exist [23, 24], similar to the one presented here. The interested reader can 
find a discussion of
these parametrizations and more comparisons with data in Ref. [25].

\vskip 5 mm
\noindent {\bf 3. \underbar{Jet production in $\gamma \gamma$ collisions and 
in photoproduction}} \par
\vskip 5 mm
The jet production in $\gamma \gamma$ collisions is a powerful tool for 
measuring parton
distributions in the photon~; the cross-section is indeed very sensitive to 
these distributions
since they may intervene twice in calculations. In this reaction, as in the 
photoproduction case, it
is important to compare the experimental results with theoretical expressions 
calculated beyond the
Leading Logarithm approximation. It  is only under these conditions that 
precise parton
distributions can be extracted from data. A detailed discussion of the jet 
production at TRISTAN has
been given in Ref. 7 that I summarize here. \par

When an inclusive jet cross-section is calculated beyond the LL aproximation, 
both the parton
distributions and the subprocesses must be corrected. For ins\-tan\-ce, let us 
consider the reaction
of Fig. 2 which can be symbolically written
$${d \sigma^{jet} \over d \vec{p}_{\bot} dy} = \sum_{ij} {\cal P}_{\gamma}^i 
\otimes \widehat{\sigma}
\left ( ij \to jet \ X \right ) \otimes {\cal P}_P^j \eqno(13)$$

\noindent where ${\cal P}_{\gamma}^i$ and ${\cal P}_P^j$ are the parton 
distributions in the photon
and in the proton, and where $\widehat{\sigma}$ is the subprocess 
cross-section. Expanding ${\cal
P}_{\gamma}^i$ and $\widehat{\sigma}$ in power of $\alpha_s$, we obtain an 
expression 
$${d \sigma^{jet} \over d \vec{p}_{\bot} \ d \eta} = \sum_{i,j} \left ( {4 \pi 
\over
\alpha_s(p_{\bot}^2)} a_i + b_i \right ) \otimes \left ( \alpha \ \alpha_s(p_{
\bot}^2)
\widehat{\sigma}_{ij}^{BORN} + \alpha \ \alpha_s^2(p_{\bot}^2) K_{ij} \right ) 
\otimes
{\cal P}_P^j \eqno(14)$$

\noindent which shows the Leading Logarithm  contributions to the jet 
cross-section
(as\-so\-cia\-ted with $a_i$ and $\widehat{\sigma}_i^{BORN}$ which describe 
the $2 \to 2$
subprocesses), and the BLL QCD corrections coming from $b_i$ and $K_{ij}$. I 
do not
discuss the well-known parton distributions in the proton and concentrate on 
the incident photon.
\par 

The term $b_i$ describes the effects of the HO corrections to the evolution 
equation (5) and it is
proportional to $k^{(1)}$, $P^{(1)}$ (cf. eq. (6)) and $\beta_1$ (the two-loop 
coefficients of the
function $\beta (\alpha_s)$). These parton distributions have been discussed 
in section~2. \par

The HO terms $K_{ij}$ correspond to $2 \to 3$ subprocesses and to virtual 
corrections to the $2 \to
2$ Born subprocesses. They contain terms which compensate the scale-dependence 
of the LL
expressions and make the corrected inclusive cross-section more stable with 
respect to variations of
the renormalization and factorization scales $\mu$ and $M$ [7]. \par

The theoretical results compared below with TRISTAN data are obtained with the 
distribution
functions described in section 2~; they include BLL QCD corrections and the 
scales have been set
equal to $p_{\bot}$~: $\mu = M = p_{\bot}$. The effects of the convolution 
with the
Weizs\"acker-Williams photon spectrum in the electron is also carefully 
discussed in Ref. [7]. 

\vbox to 11 truecm {}
 \baselineskip = 14 pt 
\noindent Fig. 8 : TOPAZ data [5] on inclusive jet production and theoretical
predictions for\break \noindent $\int_{-.7}^{.7} d \eta {d \sigma^{e^+e^- \to 
jet} \over dp_T \ d
\eta}$. The top curve is the theoretical prediction based on the standard 
photon structure functions,
the middle one is based on structure functions with half the VDM input, and 
the lower one is based on
the perturbative component only. The dash-dotted curve is the ``direct 
contribution''.\par

\vbox to 14 truecm {}
 \baselineskip = 14 pt 
\noindent Fig. 9 : AMY data [6] on inclusive jet production and theoretical
predictions for\break \noindent $\int_{-.1}^{.1} d \eta \ d \sigma/dp_{\bot} 
\ d \eta$.
The top curve is the theoretical prediction based on the standard photon 
structure functions.\par

\vskip 5 mm
\baselineskip = 20 pt 
We observe a good agreement between theory and data in which the contribution 
due to the parton
distributions in the photon play an essential role. Without the latter, the 
theoretical prediction
would be given by the dash-dotted curve of Fig. 8. We also notice that the 
theoretical cross-section
is not very sensitive to the non-perturbative inputs of the parton 
distributions~: so that we can
say that the good agreement found in Fig. 8 and 9 is a great success of 
perturbation QCD. \par

The observed disagreements for $p_{\bot}$ smaller than 5 GeV/c may be, at 
least partly, attributed
to the fact that we neglected the charm quark mass in our calculations, thus 
enhancing by some 15
$\%$ the cross section at $p_{\bot}$ = 3 GeV/c [7]. \par
Let us conclude this study of $\gamma \gamma$ collisions by saying that more 
precise data should
certainly improve the quality of the test of QCD and shall allow us to 
constrain the
non-perturbative inputs. \par

The photoproduction of large-$p_{\bot}$ jet at HERA is another source of novel 
indications
concerning the parton distributions in the photon (Fig. 2). In this case as 
well BLL calculations of
the jet cross-section are available [4, 26] and may be compared with recent 
experimental results
[27, 28]. In Fig. 10 a comparison between a theoretical prediction [8] and 
data [27] is displayed.

\vbox to 11 truecm {}
 \baselineskip = 14 pt 
\noindent  Fig. 10 : Jet cross-section as a function of the pseudo-rapidity $
\eta$~: full line.  
(The cross-section is integrated over $p_{\bot}$ from $p_{\bot}$=$7{\rm 
GeV/c}$). Gluon distribution
in the photon = 0 : dashed line. Gluon in the proton = 0 : dots.  

\vskip 5 mm
\baselineskip = 20 pt 
We notice a reasonable agreement between theory and data. Dashed and dotted 
lines in Fig. 10 show
the sensitivity of this jet cross-section to the gluon contents of the photon 
and of the proton, and
demonstrate that the rapidity dependence is a very good tool to explore and 
measure these
distributions.

\vskip 5 mm
\noindent {\bf 4. \underbar{Conclusion}} \par
\vskip 5 mm
In this talk I have presented a set of parton distributions in real photons 
which take into account
Beyond Leading Logarithm QCD corrections. I also showed how to modify the non 
perturbative input to
make it factorization scheme independent. \par

The agreement between theory and data is satisfactory as regards the photon 
structure function
$F_2^{\gamma}(x, Q^2)$. However large error bars do not allow us to draw any 
definite conclusion and
make necessary the comparison with other experimental results. \par

Particularly promising is the jet production in photon-hadron and 
photon-photon reactions. The
sensitivity of the cross-sections to the parton distributions is large and 
should allow a good
determination of the gluon contents of the photon. The good agreement between 
theory and data in
what concerns the jet cross-section $d \sigma (\gamma \gamma \to jet + X)/d
\vec{p}_{\bot} \ d\eta$
is a success for perturbative QCD.
\vskip 5 mm
\noindent {\bf \underbar{Acknowledgements}} \par
I would like to thank Dr. T. Tauchi and his colleagues for the organization of 
this interesting
meeting, and Dr. Y. Shimizu and the ``Minami Tateya'' group for their warm 
hospitality and for many
interesting discussions.
 \vfill \supereject
\centerline{\bf \underbar{References}} \par \bigskip
\item{[1]} Reviews on the photon structure function may be found
in C. Berger and W. Wagner, Phys. Rep. $\underline{146}$ (1987) 1 ; H.
Kolanoski and P. Zerwas, in High Energy ${\rm e}^+$-${\rm e}^-$ physics, World
Scientific, Singapore, 1988, Eds. A. Ali and P. ${\rm S\ddot o ding}$.
\item{} J. H. Da Luz Vieira and J. K. Storrow, Z. Phys. $\underline{C51}$
(1991) 241.  
\item{[2]} H1 Collaboration, T. Ahmed et al., Phys. Lett. $\underline{B297}$
(1992) 205.  
\item{[3]} Zeus Collaboration, M. Derrick et al., Phys. Lett.
$\underline{B297}$ (1992) 404.  
\item{[4]} For a review on the photoproduction at HERA, see M. Fontannaz,
``The Photon Structure Function at HERA", Orsay preprint LPTHE 93-22, talk
given at the XXI International Meeting on Fundamental Physics, Miraflores
de la Sierra, Spain (May 1993), and ``Hard Processes and Polarization 
Phenomena in QCD'', talk
at the ``Workshop on Two-Photon Physics at LEP and HERA'', Lund 1994, edited 
by G. Jarlskog and
L. J\"onsson, Lund University.
\item{[5]} TOPAZ collaboration, H. Hayashi et al., Phys. Lett. \underbar{B314} 
(1993) 149.
\item {[6]} AMY collaboration, R. Tomaka et al., Phys. Lett.
\underbar{B325} (1994) 248. \item{[7]} P. Aurenche, M. Fontannaz, J. Ph. 
Guillet,  Y. Shimizu, J.
Fujimoto and K. Kato, Prog. Theor. Physics \underbar{92} (1994) 175~; \item{} 
P. Aurenche et al.,
talk at the ``Workshop on Two-Photon Physics at LEP and HERA'', Lund 1994, 
edited by G. Jarlskog and
L. J\"onsson, Lund University. \item {[8]} P. Aurenche, M. Fontannaz and J. 
Ph. Guillet, Phys. Lett.
\underbar{B338} (1994) 98.  \item{[9]} I. Antoniadis and G. Grunberg, Nucl. 
Phys. \underbar{B213}
(1983) 455. \item{[10]} A. S. Gorski, B. L. Ioffe, A. Yu Khodjaminian, A. 
Oganesian, Z. Phys.
\underbar{C44} (1989) 523.
 \item{[11]} E. Witten, Nucl. Phys.  \underbar{B210} (1977) 189.
\item{[12]} A review on QCD corrections, and previous references on the 
subject can be found in~: F.
M. Borzumati and G. A. Schuler, Z. Phys. \underbar{C58} (1993) 139. 
\item{[13]} M. Gl\"uck and E. Reya, Phys. Rev. \underbar{D28} (1983) 2749.
\item{[14]} W. A. Bardeen and A. J. Buras, Phys. Rev. \underbar{D20} (1979) 
166 ; \underbar{21}
(1980) 2041 (E).
 \item{[15]} M. Fontannaz and E. Pilon, Phys. Rev. \underbar{D45} (1992) 382.
\item{[16]} E. Laenen, S. Riemersma, J. Smith and W. L. van Neerven, Phys. 
Rev. \underbar{D49}
(1994) 5753. 
\item{[17]} M. Gl\"uck, E. Reya and A. Vogt, Phys. Rev. \underbar{D45} (1992) 
3986.
\item{[18]} P. Aurenche, M. Fontannaz and J.-Ph. Guillet, Z. Phys. C. (1994).
\item{[19]} T. Sasaki et al., Phys. Lett. \underbar{B252} (1990) 491.
\item{[20]} JADE collaboration, W. Bartel et al., Z. Phys. \underbar{C24} 
(1984) 231.
\item{[21]} PLUTO collaboration, Ch. Berger et al., Nucl. Phys. 
\underbar{B281} (1987) 365~; Phys.
Lett. \underbar{142B} (1984) 11.
\item{[22]} TOPAZ collaboration, K. Muramatsu et al., Phys. Lett. 
\underbar{B332} (1994) 447.
 \item{[23]} M. Gl\"uck, E. Reya and A. Vogt, Phys. Rev. \underbar{D46} (1992) 
1973. 
\item{[24]} L. E.
Gordon and J. K. Storrow, Z. Phys. \underbar{C56} (1992) 307. 
\item{[25]} A. Vogt, Proceedings of the
Workshop ``Two-Photon Physics at LEP and HERA'', Lund, May 1994. 
\item{[26]} J. Ph. Guillet, talk at the ``Workshop on Two-Photon Physics at 
LEP and HERA'', Lund
1994.  
\item {[27]} H1 collaboration, ``27th International Conference on HEP'', 
Glasgow 1994, presented
by H. Hufnagel. 
\item {[28]} ZEUS collaboration, M. Derrick et al., DESY 94-176.

 \bye